\documentclass[prl,showpacs,twocolumn,superscriptaddress]{revtex4}
\usepackage{amsmath,amssymb,amsfonts}
\usepackage{amsthm}
\usepackage{graphicx}
\usepackage{pmat}

\newcommand{\bra}[1]{\ensuremath{\langle #1 |}}
\newcommand{\ket}[1]{\ensuremath{| #1 \rangle}}

\newcommand{\bk}[2]{\langle #1|#2\rangle}
\newcommand{\kb}[1]{| #1  \rangle\langle #1|}

\DeclareMathOperator{\tr}{tr}

\def\one{\leavevmode\hbox{\small1\normalsize\kern-.33em1}}
\newcommand{\CC}{\mathbb{C}}
\newcommand{\RR}{\mathbb{R}}

\newcommand{\BB}[1]{\mathcal{B}_\mathrm{#1}}
\newcommand{\YY}[2]{\left\langle {{Y}}_\mathrm{#1} {{Y}}_\mathrm{#2} \right\rangle}

\newcommand{\AB}{\langle \mathcal{B}_{\text{AB}}\rangle}
\newcommand{\AC}{\langle \mathcal{B}_{\text{AC}}\rangle}
\newcommand{\BC}{\langle \mathcal{B}_{\text{BC}}\rangle}

\def\PP{{P}}

\newtheorem{theorem}{Theorem}
\newtheorem{lemma}[theorem]{Lemma}

\theoremstyle{definition}

\theoremstyle{remark}

\begin{document}
\title{Monogamy of Bell correlations and \uppercase{T}sirelson's bound}
\author{Benjamin F. Toner}
\email{Ben.Toner@cwi.nl}
\affiliation{Institute for Quantum Information, California Institute of
Technology, Pasadena, CA 91125, USA  }
\affiliation{Centrum voor Wiskunde en Informatica, Kruislaan 413, 1098 SJ Amsterdam, The Netherlands}
\date{\today}
\author{Frank Verstraete}
\email{frank.verstraete@univie.ac.at}
 \affiliation{Institute for Quantum Information, California
Institute of Technology, Pasadena, CA 91125, USA  } \affiliation{Fakult\"at f\"ur Physik, Universit\"at Wien,
Austria}
\date{\today}


\begin{abstract}
  We consider three parties, A, B, and C, each performing one of two
  local measurements on a shared quantum state of arbitrary dimension.
  We characterize the trade-off between the nonlocality of the Bell
  correlations observed by AB and of those observed by AC. This
  generalizes Tsirelson's bound on the quantum value of the CHSH
  inequality, the latter being recovered when C is completely
  uncorrelated with AB. We also discuss the trade-off between Bell
  violations and local expectation values of observables that
  anticommute with the ones used in the Bell test.
\end{abstract}


\maketitle

The existence of Bell inequalities~\cite{Bell:64a,Clauser:69a} and
their observed violation in experiments has had a very deep impact on
the way we look at quantum mechanics. On the one hand, it has led to a
study of the precise meaning of nonlocality and opened up the field of
entanglement theory. On the other, it has led to the observation that
Bell violations can be exploited in the design of cryptographic
protocols~\cite{PhysRevLett.67.661}.
In that case, an eavesdropper (C) tries to gain access to some quantum
correlations shared by Alice (A) and Bob (B). If the Bell correlations
between A and B are strong, it can happen that C's outcomes will be
almost uncorrelated with them, and A and B will be able to execute a
purification protocol so as to create private randomness. In the
present paper, we will make a precise quantitative statement about the
following monogamy property: Suppose A and B violate a Bell inequality
by a certain amount.  How does that bound the possible Bell
correlations between A and C?  This is also interesting from the point
of view of entanglement theory, as it provides monogamy relations
independent of the size of the local Hilbert spaces. For the
Clauser-Horne-Shimony-Holt (CHSH) inequality, the region of accessible
Bell correlations between AB and AC turns out to be very simple (see Figure~1).

\begin{figure}[b]  \centering
\includegraphics{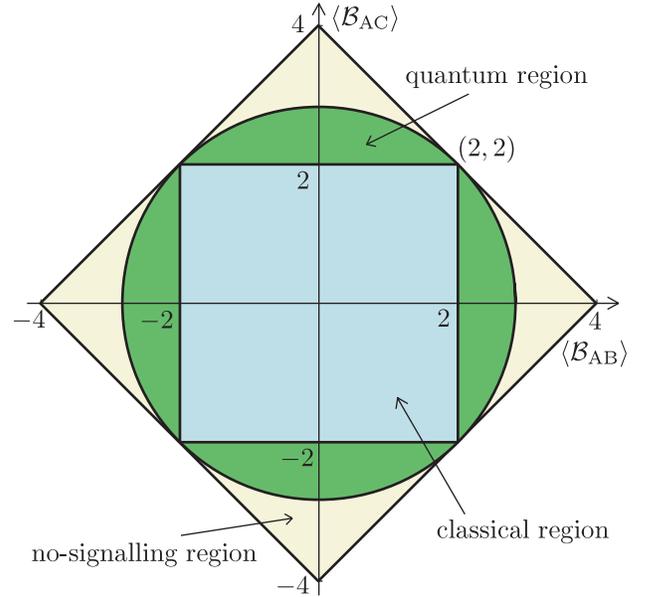}
\caption{Accessible values of $\AB$ and $\AC$ for classical theories (interior of square), quantum theory (interior of
circle), and no-signalling theories (interior of diamond). Note that both quantum and no-signalling theories obey
monogamy constraints; classical local hidden variable theories do not.} \label{fig:1}
\end{figure}

In the setting where two parties, A and B, share a quantum state $\rho$, and each has the choice of two local
measurements, there is just one relevant Bell inequality, the CHSH
inequality~\cite{Clauser:69a}.  Define the {CHSH operator}
\begin{equation}
\label{eq:1}
  {\cal B}_\text{AB} = {A}_1 \otimes \left({B}_1 + {B}_2\right) +
  {A}_2 \otimes \left({B}_1 - {B}_2\right),
\end{equation}
where ${A}_1$ and ${A}_2$ (${B}_1$ and ${B}_2$) are A's (B's) observables and are Hermitian operators with
spectrum in $[-1,+1]$.  For particular measurements and a particular state $\rho$, the quantum value of the CHSH
inequality is defined as $\AB = \tr \left({\cal B}_\text{AB} \rho\right)$.  All correlations described by local
hidden variable (LHV) models satisfy the CHSH inequality, $\left|\AB_\text{LHV} \right| \leq 2 $, but in the case
of entangled quantum systems, this bound can be violated.  For example, on the singlet state of two qubits there exist
operators $A_i,B_i$ such that $\AB = \bra{\psi^-} {\cal B}_\text{AB}\ket{\psi^-} = 2\sqrt2 $.

We do not yet know how to calculate a bound on the maximum quantum value of an arbitrary Bell inequality, but a
number of ad hoc techniques have been
developed~\cite{Tsirelson:80a,Tsirelson:85b,toner:_monog,navascues:_bound}.
In the case of the CHSH inequality, Tsirelson has proved that $ \left|\tr \left( {\cal B}_\text{AB}\rho\right)
\right| \leq 2\sqrt2 $ for all observables ${A}_1$, ${A}_2$, ${B}_1$, ${B}_2$, and all states
$\rho$~\cite{Tsirelson:80a}.  This Tsirelson bound can itself be violated if we consider more general hypothetical
{\em no-signalling} theories: a {\em nonlocal box} violates the CHSH inequality maximally, $\AB_\text{NL} =
4$~\cite{Popescu:94a}. Tsirelson's bound is a simple mathematical consequence of the axioms of
quantum theory, but is there some deeper reason why a violation greater than $2 \sqrt 2$ is unphysical?  For
example, a violation greater than $\sqrt{32/3}\approx 3.27$ would imply that any communication complexity problem
can be solved using a constant amount of communication~\cite{dam05:_implaus_conseq_super_nonloc}.
As will be clear from the following results, the bound $2\sqrt{2}$ is very natural if one considers the possible
Bell violations in a three-party setup.

We establish the following monogamy trade-off:
\begin{theorem}
\label{thm:1}
Suppose that three parties, A, B, and C, share a quantum
state (of arbitrary dimension) and each chooses to measure one of two
observables.  Then
\begin{equation}
\label{eq:2}
\AB^2 + \AC ^2 \leq 8.
\end{equation}
\end{theorem}
Here,  $\mathcal{B}_{\text{AC}}$ is defined as in Eq.~(\ref{eq:1}), but with B's observables replaced by C's. Note
that we obtain Tsirelson's bound, $\AB^2 \leq 8$, as a simple corollary.  Note also that A's measurements are the
same in $\AB$ and $\AC$: otherwise we could have $\AB = \AC = 2\sqrt 2$ and there would be no trade-off.
Theorem~\ref{thm:1} is analogous to the Coffman-Kundu-Wootters theorem that describes the trade-off between how
entangled A is with B, and how entangled A is with C~\cite{Coffman:00a}. Eq.~(\ref{eq:2}) is the best
possible bound: there are states and measurements achieving any values of $\AB$ and $\AC$ that satisfy it.
Previously the best bound known was
$
\left|\AB\right| + \left|\AC\right| \leq 4$,
which is tight for correlations that arise from no-signalling theories~\cite{toner:_monog,masanes05:_gener_nonsig_theor}.
We illustrate the monogamy trade-offs for various theories in Figure~\ref{fig:1}.

We prove Theorem~\ref{thm:1} in two parts.  We first show that is
sufficient to restrict to states with support on a qubit at each site.
We can then relax the requirement that A's measurements be the same in
$\AB$ and $\AC$, maximizing over the measurements in $\AB$ and $\AC$
separately, but keeping the state fixed.  Our proof suggests a
connection between anticommutation and Bell inequality violation,
which we then explore more deeply.

{\em Dimensional reduction}.---We start by establishing a bound on the dimension of the quantum state  required to
maximally violate certain Bell inequalities.  A similar result was proved by Masanes~\cite{masanes05:_extrem}. The
main ingredient---a canonical decomposition for a pair of subspaces of $\CC^n$---is described in more detail in,
e.g., Ref.~\cite{bhatia96:_matrix_analy}.

\begin{lemma} \label{lemma:circle1}
Consider any Bell inequality in the setting where $m$
parties each choose from two two-outcome measurements.  Then the
maximum quantum value of the Bell inequality is achieved by a state
that has support on a qubit at each site.  Furthermore, we can assume
this state has real coefficients and that the observables are real
and traceless.
\end{lemma}

\begin{proof}
 For $i \in \{1,2\}$, assume party $k$ has observables ${M}_{k,i}$,
acting on a Hilbert space ${\cal H}_k$.  By
extending the local Hilbert spaces ${\cal H}_k$, we can assume for all
$k$ and for all $i = 1,2$ that (i) ${\cal H}_k = \CC^{2d}$ for some
fixed $d$, (ii)
${M}_{k,i}$ has eigenvalues $\pm 1$, and (iii) $\tr {M}_{k,i}
= 0$.  The first condition states that all local spaces have the same
dimension $2d$, the latter two that each observable corresponds to
a projective measurement onto a $d$-dimensional subspace and its
complement.  We also define ${  M}_{k,0} = \one_{2d}$, the identity operator on ${\cal H}_k$.  We can
write a generic Bell operator in the setting stated in the lemma as
\begin{equation}
{\cal B} = \sum_{i_1=0}^2 \sum_{i_2=0}^2 \cdots \sum_{i_m=0}^2 c_{i_1i_2\cdots i_m}\bigotimes_{k=1}^m {M}_{k,i_k},
\label{circle-eq:3}
\end{equation}
where the coefficients $c_{i_1i_2\cdots i_m}$ are arbitrary real numbers.
Our goal is find the quantum value of this Bell operator, which is
maximum of $B \equiv \bra \psi \cal B \ket \psi$ over states $\ket \psi$ and
measurements ${M}_{k,i}$.

We now choose a local basis for each ${\cal H}_k$ such that party $k$'s
observables have a simple form.  We start by taking
${M}_{k,1} = \begin{pmat}[{|}]\one_d & {0} \cr\- {0}
    & -\one_d \cr
  \end{pmat}$.  This leaves us the freedom to specify the basis within
  the two $d\times d$ blocks on which ${M}_{k,1}$ is constant.  Let
  ${M}_{k,2} = 2 \PP \PP^\dagger - \one_{2d}$ (we suppress the
  dependence on $k$), where $\PP$ is a $2d \times d$ matrix with
  orthonormal columns, which span the $+1$--eigenspace of
  ${M}_{k,2}$).  Write $\PP = \begin{pmat}[{}]\PP_1 \cr \- \PP_2
    \cr\end{pmat}$, where $\PP_1$ and $\PP_2$ are $d \times d$
  matrices.  The rows of $P$ are orthonormal, which implies ${
    P}^\dagger {P} = { P}_1^\dagger {P}_1 + { P}_2^\dagger {P}_2 =
  \one_d$, so $ { P}_1^\dagger {P}_1$ and ${ P}_2^\dagger {P}_2$ are
  simultaneously diagonalizable.  This means there is a singular
  value decomposition of the form $ \PP_1 = {U}_1^\dagger {D}_1{V}$,
  $\PP_2 = {U}_2^\dagger {D}_2{V}$, where ${U}_1$, ${U}_2$ and $V$ are
  $d\times d$ unitary matrices and ${D}_1$ and ${D}_2 = \sqrt{\one_d -
    {D}_1^2}$ are nonnegative (real) diagonal matrices.
Changing basis according to
  the unitary ${U}_1 \oplus {U}_2$, which leaves ${M}_{k,1}$ invariant, it
  follows that
${M}_{k,2}
  = \begin{pmat}[{|}]2{D}_1^2 - \one_d & 2 {D}_1 {D}_2\cr \-
   2 {D}_1 {D}_2& 2{D}_2^2 - \one_d \cr \end{pmat}$,
 where each of
  the $d \times d$ blocks is diagonal.  We relabel
  our basis vectors so that ${M}_{k,1} = \bigoplus_{j=1}^d {Z}$, $M_{k,2} =
  \bigoplus_{j=1}^d \left(\cos \theta_j {Z}  + \sin \theta_j
  {X} \right)$, where $2{D}_1^2 - \one_d
  = \mathrm{diag}(\cos \theta_1, \cos \theta_2, \ldots, \cos
  \theta_d)$ and $X$ and $Z$ are the usual Pauli operators.  Hence our operators are real and preserve a $\oplus_{j = 1}^d \CC^2$
  subspace of ${\cal H}_k$.  They are traceless on each $\CC^2$ space.

We wish to maximize $B = \bra \psi \cal B \ket \psi$ over the state
$\ket \psi$ and the
measurements ${M}_{k,i}$.  Fix $k$, and let $\rho_{k,j}$ be the reduced density
matrix obtained by projecting $\ket \psi$ onto the $j$'th $\CC^2$
factor of the $\oplus_{j = 1}^d \CC^2$ subspace induced by
$M_{k,1}$ and $M_{k,2}$ at site $k$.
Then $B = \sum_{j=1}^d \tr
{\cal B} \rho_{k,j}$ is a convex sum over the $\CC^2$ factors, whereupon it follows that the maximum is
achieved by a state with support on a qubit at site $k$.  Since this
argument works for all $k$, the maximum of $B$ is
achieved by a state that has support on a qubit on each site.

Finally, write $\ket \psi = \ket {\psi_1} + i \ket {\psi_2}$, where
$\ket {\psi_1}$ and $\ket {\psi_2}$ are real.  Then $\bra \psi {\cal
  B} \ket \psi = \bra {\psi_1} {\cal B} \ket {\psi_1} + \bra {\psi_2}
{\cal B} \ket {\psi_2}$ since $\cal B$ is real, which is the same
expression we would obtain if the state were a real mixture of $\ket
{\psi_1}$ and $\ket {\psi_2}$.  Hence the maximum of $B$ is achieved
by a state with real coefficients.
\end{proof}

{\em Monogamy trade-off relation.}---The region $\cal R$ of allowed
values of $\left(\AB, \AC \right)$ is convex and can therefore be
described by an (infinite) family of half-space inequalities,
\begin{equation}
c_{\text{AB}}\AB + c_{\text{AC}}\AC  \leq d,
\label{circle-eq:4}
\end{equation}
with $c_{\text{AB}}, c_{\text{AC}}, d \in \RR$.  The left-hand side of
Eq.~(\ref{circle-eq:4}) is a Bell operator, as defined in Eq.~(\ref{circle-eq:3}),
which means we can apply Lemma~\ref{lemma:circle1} to conclude that extreme points of
$\cal R$ are achieved by real states on three qubits, with measurements of the form ${M} = \cos \theta {Z} +
\sin \theta {X}$.  Theorem~1 will emerge as a corollary of:
\begin{lemma}
\label{lemma:circle2}
Let $\ket \psi$ be a pure state in $\CC^2 \otimes \CC^2 \otimes
\CC^2$ with real coefficients.  Then the maximum of $\left\langle\BB{AB}\right\rangle$ over
real traceless observables ${A}_1, {A}_2, {B}_1, { B}_2$ is
\begin{equation}
  \label{circle-eq:8}
 2 \sqrt {1 + \YY{A}B^2-\YY{A}C^2-\YY{B}C^2},
\end{equation}
where $Y$ is the usual Pauli operator, $\YY{A}B = \tr \left(Y_A
  \otimes Y_B \otimes \one\, \rho\right)$, and so on.  Cyclic
permutations of Eq.~(\ref{circle-eq:8}) hold for $\langle\BB{AC}\rangle$ and
$\langle\BB{BC}\rangle$.
\end{lemma}

\begin{proof}We consider $\rho_{AB} = \tr_C \kb \psi$, which is a real
state on $\CC^2 \otimes \CC^2$.  Horodecki and family have calculated the maximum quantum value of the CHSH
operator for a state on $\CC^2 \otimes \CC^2$~\cite{horodecki96:_separ}.  Their analysis simplifies in our case
because the state and measurements are real.  Define
\begin{equation}
  \label{eq:12}
T_\text{AB} = \left[\begin{matrix}\langle X_A X_B\rangle & \langle X_A Z_B\rangle \\
\langle Z_A X_B\rangle & \langle Z_A Z_B\rangle
\end{matrix}\right].
\end{equation}
For $i=1,2$, write ${A}_i = \hat {a}_i \cdot \vec {\sigma}_r$, ${B}_i = {\hat b}_i \cdot \vec \sigma_r$, where
$\hat a_i$ and $\hat b_i$ are two-dimensional unit vectors and $\vec \sigma_r = ({X}, {Z})$.  Define
\begin{equation}
  \label{circle-eq:5}
  \hat {b}_1 + \hat {b}_2 = 2 \cos \theta \hat {d}_1,  \ \   \hat {b}_1 - \hat {b}_2 = 2 \sin \theta \hat {d}_2,
\end{equation}
where $\theta \in [0, \pi/2]$ and $\hat {d}_1$ and $\hat {d}_1$ are orthogonal unit vectors.  
Then
  \begin{eqnarray}
\frac12 \max_{{A}_i, {B}_j} \left \langle\BB{AB}\right \rangle &=& \max_{\hat {d}_i, \theta, \hat {a}_i} \cos \theta \hat {a}_1^t T_\text{AB} \hat {d}_1 + \sin \theta \hat {a}_2^t T_\text{AB} \hat {d}_2    \nonumber \\
 &=& \max_{\hat {d}_i, \theta} \cos \theta \left\|T_\text{AB} \hat {d}_1\right\| + \sin \theta \left\| T_\text{AB} \hat {d}_2\right\| \nonumber \\
&=& \max_{\hat {d}_i} \sqrt{\left\|T_\text{AB} \hat {d}_1\right\|^2 + \left\| T_\text{AB} \hat {d}_2\right\|^2}\label{eq:d}\\
&=& \sqrt {\tr \left({T_\text{AB}} {T_\text{AB}}^t\right)}.
\end{eqnarray}
This is just the Frobenius norm of $T_\text{AB}$ and it is straightforward to check that, for pure states on
$\CC^2 \otimes \CC^2 \otimes \CC^2$ with real coefficients, it is equal to half of
Eq.~(\ref{circle-eq:8}).\end{proof}
\begin{lemma}
\label{sec:dimens-reduct-1}
For a pure state $\ket \psi$ with real coefficients in $\CC^2 \otimes \CC^2 \otimes \CC^2$,
\begin{equation}
\label{eq:9}
\max_{{A}_i, {B}_j, {C}_k} \left \langle \BB{AB}\right \rangle^2 + \left \langle \BB{AC}\right\rangle^2 = 8 \left(1 - \YY{B}C^2 \right).
\end{equation}
\end{lemma}
\begin{proof}
Lemma~\ref{lemma:circle2}, applied to $\AB$ and $\AC$ separately, immediately implies:
\begin{eqnarray}
\label{eq:10}
\max_{{A}_i, {B}_j, {C}_k} \left \langle \BB{AB}\right \rangle^2 + \left \langle \BB{AC}\right\rangle^2 &\leq&
\max_{{A}_i, {B}_j} \left \langle \BB{AB}\right \rangle^2 + \max_{{A}_i, {C}_k} \left \langle \BB{AC}\right\rangle^2\nonumber \\
 &=& 8 \left(1 - \YY{B}C^2 \right).
\end{eqnarray}
The reason we do not have equality is that the measurements $A_i$ achieving the maximum in $\AB$ and $\AC$ may be different.  We have to show they can be chosen to be the same.  Define $T_\text{AC}$ in analogy with Eq.~(\ref{eq:12}) and write the vectors corresponding to C's measurements as
\begin{equation}
\label{eq:13}
  \hat {c}_1 + \hat {c}_2 = 2 \cos \theta \hat {e}_1,  \ \   \hat {c}_1 - \hat {c}_2 = 2 \sin \theta \hat {e}_2,
\end{equation}
in analogy with Eq.~(\ref{circle-eq:5}) for B's observables.  One can check that $\left[ T_\text{AB} T_\text{AB}^t, T_\text{AC} T_\text{AC}^t\right] = 0$ for all pure states $\ket \psi$ with real coefficients.  Hence there are orthonormal vectors $a'_1$ and $a'_2$ that are simultaneous eigenvectors of $T_\text{AB} T_\text{AB}^t$ and $T_\text{AC} T_\text{AC}^t$.  Next, note that the term being maximized in Eq.~(\ref{eq:d}), $\|T_\text{AB} \hat d_1
\|^2 + \|T_\text{AB} \hat d_2 \|^2$, is actually independent of the
$\hat d_i$ (recall that $\hat d_1 \cdot \hat d_2 =0$), so we are free to choose the $\hat d_i$ as we please.  Take $\hat d_i = T^t_\text{AB} \hat a_i'$ for $i = 1, 2$ and, similarly, take $\hat e_i = T^t_\text{AC} \hat a_i'$.  Alice's measurement vector $\hat a_i$ in the AB maximization of the previous lemma was taken to be the unit vector along $T_\text{AB} \hat d_i$, but this is $T_\text{AB} T^t_\text{AB} \hat a'_i \propto \hat a'_i$ so $\hat a_i = \hat a'_i$.  The same will hold in the AC maximization.  Hence we can choose A's measurement vectors to be the same in both cases, and we have equality in Eq.~(\ref{eq:9}).
\end{proof}

Combining Lemmas~\ref{lemma:circle1} and~\ref{sec:dimens-reduct-1}, we obtain Theorem~\ref{thm:1}.

{\em The monogamy trade-off is tight}.---Lemma~\ref{sec:dimens-reduct-1} also implies that any $\AB$ and $\AC$ compatible with Eq.~(\ref{eq:9}) are achievable.  In particular,
  the state
\begin{equation}
  \label{eq:14}
  \ket \psi = c_- \left(\ket {010} + \ket {011} \right) + c_+ \left( \ket {100} + \ket {101}\right),
\end{equation}
where
\begin{equation}
  \label{eq:15}
c_\pm = \frac12 \, \sqrt{1 \pm \sqrt 2 \sin t},
\end{equation}
and $0 \leq t \leq \pi/4$, gives $\AB = 2 \sqrt2 \, \cos t$, $\AC = 2 \sqrt2 \, \sin t$.

{\em Extensions.}---In the case of the CHSH inequality we can, in principle, obtain monogamy trade-offs
when there are more than three parties via Lemma~\ref{lemma:circle1}, which converts the problem into a finite
optimization problem.  In the three-party setting, if we are interested in $\BC$ as well as $\AB$ and $\AC$ then
we can obtain the trade-off surface numerically.  The technique of Lemma~\ref{sec:dimens-reduct-1}---to allow A's
measurements to be different in $\AB$ and $\AC$ and then show that they could be the same anyway---does not work.
It predicts that the trade-off surface be the intersection of the three cylinders, $\AB^2 + \AC^2 \leq 8$, $\AB^2 +
\BC^2 \leq 8$, and $\AC^2 + \BC^2 \leq 8$, but one can show, for example by using the multipartite generalization of
Navascues, Pironio and Ac{\'\i}n's semidefinite programming bounds~\cite{navascues:_bound}, that there are points on
this surface that are not achievable.  It would be interesting to extend the semidefinite programming technique to
obtain monogamy inequalities for other Bell inequalities.



{\em Bell inequality violation and anticommutation}.---The precise form of Eq.~(\ref{eq:9}) suggests a
general connection between the trade-off  of Bell inequality violation and the expectation values of
anticommuting observables; indeed, the operator $Y_BY_C$ anticommutes with all observables $B_j,C_k$ measured in
the Bell test. We now give a more general result, restricting for simplicity to the two party case.
\begin{theorem}
\label{thm:n}
  Let ${\cal B} = \sum_{i,j=1}^n p_{ij} A_i \otimes B_j$ be a 2-party 2-outcome correlation Bell operator such
  that $\tr \rho {\cal B}  \leq {\cal Q}$ for all shared states $\rho$ and all observables $A_i$, $B_j$ (with spectrum in $[-1,+1]$). Let $W$ be any observable (with spectrum in $[-1,+1]$) on Bob's Hilbert space that
  anticommutes with $B_j$ for all $j$. Then
\begin{equation}
\tr \rho {\cal B} \leq {\cal Q}\sqrt {1 - \left( \tr \rho W\right)^2 }.
\end{equation}
\end{theorem}
\begin{proof}
  We start by noting that it is sufficient to restrict to the case
  where $\rho$ is pure, i.e., $\rho = \ket \psi \bra \psi$.  The general case then follows by applying
  Jensen's inequality to the concave function $f(x) = \sqrt{1-x^2}$.  

  If $\ket w$ is an eigenvector of $W$ with eigenvalue $w$, $B_j\ket
  w$ is either 0 or an eigenvector of $W$ with eigenvalue $-w$, since
  $B_j$ and $W$ anticommute.  This means that there is a decomposition
  ${\cal H}_B = {\cal W}_0 \oplus {\cal W}_1 \oplus {\cal W}_2$, where
  $W$ annihilates vectors in ${\cal W}_0$, every $B_j$ annihilates
  vectors in ${\cal W}_1$, and ${\cal W}_2$ is spanned by eigenvectors of
  $W$ with nonzero eigenvalues, which occur in positive/negative pairs.

Denote the distinct positive
  eigenvalues associated with eigenvectors of $W$ in ${\cal W}_2$ as 
$0 < w_2 < w_3 < \cdots < w_m \leq 1$ and
  let $V_i^\pm$ be the subspace in ${\cal W}_2$ corresponding to eigenvalue
  $\pm w_i$.  Decompose $\ket \psi$ as $\ket \psi = \sum_{i=0}^m\sqrt{p_i}\ket {\psi_i}$, where $\sum_i p_i = 1$, $\ket {\psi_0} \in {\cal W}_0$, $\ket {\psi_1} \in {\cal W}_1$, $\bk {\psi_0} {\psi_0} = \bk {\psi_1}{\psi_1} = 1$, and for $i \geq 2$, $\ket {\psi_i} = \left( \cos \theta_i \ket{w_i^+} +
    \sin \theta_i\ket{w_i^-}\right)$, $\ket {w_i^\pm} \in V_i^\pm$, and 
  $\bk {w_i^\pm}{w_i^\pm}=1$.  Then
${\cal B}\ket {w_i^\pm} \in V_i^\mp$ for $i \geq 2$. 
It follows that
\begin{eqnarray}
\bra \psi {\cal B} \ket\psi &=& p_0 \bra {\psi_0} {\cal B} \ket {\psi_0} + \sum_{i=2}^m  p_i \mathrm{Re}\left(\bra {w_i^+} {\cal B} \ket {w_i^-} \right)\sin 2 \theta_i,\nonumber\\
\bra \psi W \ket\psi  &=& p_1 \bra {\psi_1}  W \ket {\psi_1} + \sum_{i=2}^m p_i w_i^2 \cos 2 \theta_i.
\end{eqnarray}
But these are the same expressions we would obtain with the mixed state $\rho = \sum_i p_i \ket {\psi_i} \bra {\psi_i}$.  It therefore follows from our initial remark that it is sufficient to prove the claim for each state $\ket {\psi_i}$.  For $i=0,1$, the result is trivial.  Fix $i\geq 2$, set $\chi = \mathrm{Re}\left(\bra {w_i^+} {\cal B} \ket {w_i^-} \right)$ and let $x \in \{\pm1\}$ be the sign of $\chi$.  Set $\ket \phi =
\frac{1}{\sqrt2} \ket {w_i^-} + x \frac{1}{\sqrt2} \ket {w_i^+}$.
Then $\bra \phi {\cal B} \ket\phi = \chi < Q$ by assumption, while
\begin{eqnarray}
  \label{eq:3}
\bra {\psi_i} {\cal B} \ket{\psi_i} &\leq& \chi \sin  2 \theta_i\\
&\leq& \chi \sqrt{1 - \bra {\psi_i} W \ket{\psi_i}^2 }\\
&\leq& Q \sqrt{1 - \bra {\psi_i} W \ket{\psi_i}^2 }.
\end{eqnarray}
This completes the proof.
\end{proof}

We now apply Theorem~\ref{thm:n} to the CHSH inequality.  If $B_1$ and $B_2$ are
observables with $\pm 1$ eigenvalues, then they both anticommute with
their commutator $i [ B_1, B_2]/2$ (the factor of $i/2$ makes this an
observable).  Applying Theorem~\ref{thm:n} to both Alice and Bob's
observables, it follows that
\begin{eqnarray}
  \label{eq:8}
  \AB &\leq& 2 \sqrt {2 - \left|\langle [B_1, B_2] \rangle\right|^2 }. \\
  \AB &\leq& 2 \sqrt {2 - \left|\langle [A_1, A_2] \rangle\right|^2 }.
\end{eqnarray}
These are local analogues of Tsirelson's bound~\cite{Tsirelson:80a},
\begin{equation}
  \AB \leq \sqrt {4 +  \left|\langle [A_1,A_2][B_1, B_2] \rangle\right| }.\\
\end{equation}
In particular, for maximal quantum violation of the CHSH inequality, the local observables corresponding to the
commutators must be locally random $\langle [A_1, A_2] \rangle = \langle [B_1, B_2] \rangle = 0$ but perfectly
correlated $\left|\langle [A_1,A_2][B_1, B_2] \rangle\right| = 4$.  This is a clear manifestation of the fact
that entanglement goes hand in hand with local randomness.

In conclusion, we investigated Bell correlations in a tripartite setting and obtained tight monogamy bounds on
the trade-off between them. The main message is depicted in Figure~1, which gives a universal picture of nonlocal
correlations valid for quantum systems of any dimension.

{\em Acknowledgments}.---We thank Richard Cleve, Wim van Dam, Andrew Doherty, Nicolas Gisin, and Oded Regev for discussions. This
work was supported in part by the NSF under Grants PHY-0456720 and CCF-0524828, the ARO under Grant W911NF-05-1-0294, EU project QAP 015848, and NWO VICI project 639-023-302.


\begin{thebibliography}{24}
\expandafter\ifx\csname natexlab\endcsname\relax\def\natexlab#1{#1}\fi
\expandafter\ifx\csname bibnamefont\endcsname\relax
  \def\bibnamefont#1{#1}\fi
\expandafter\ifx\csname bibfnamefont\endcsname\relax
  \def\bibfnamefont#1{#1}\fi
\expandafter\ifx\csname citenamefont\endcsname\relax
  \def\citenamefont#1{#1}\fi
\expandafter\ifx\csname url\endcsname\relax
  \def\url#1{\texttt{#1}}\fi
\expandafter\ifx\csname urlprefix\endcsname\relax\def\urlprefix{URL }\fi
\providecommand{\bibinfo}[2]{#2}
\providecommand{\eprint}[2][]{\url{#2}}

\bibitem[{\citenamefont{Bell}(1964)}]{Bell:64a}
\bibinfo{author}{\bibfnamefont{J.~S.} \bibnamefont{Bell}},
  \bibinfo{journal}{Physics} \textbf{\bibinfo{volume}{1}}, \bibinfo{pages}{195}
  (\bibinfo{year}{1964}).

\bibitem[{\citenamefont{Clauser et~al.}(1969)\citenamefont{Clauser, Horne,
  Shimony, and Holt}}]{Clauser:69a}
\bibinfo{author}{\bibfnamefont{J.~F.} \bibnamefont{Clauser}},
  \bibinfo{author}{\bibfnamefont{M.~A.} \bibnamefont{Horne}},
  \bibinfo{author}{\bibfnamefont{A.}~\bibnamefont{Shimony}}, \bibnamefont{and}
  \bibinfo{author}{\bibfnamefont{R.~A.} \bibnamefont{Holt}},
  \bibinfo{journal}{Phys. Rev. Lett.} \textbf{\bibinfo{volume}{23}},
  \bibinfo{pages}{880} (\bibinfo{year}{1969}).

\bibitem[{\citenamefont{Ekert}(1991)}]{PhysRevLett.67.661}
\bibinfo{author}{\bibfnamefont{A.~K.} \bibnamefont{Ekert}},
  \bibinfo{journal}{Phys. Rev. Lett.} \textbf{\bibinfo{volume}{67}},
  \bibinfo{pages}{661} (\bibinfo{year}{1991});
\bibinfo{author}{\bibfnamefont{C.~H.} \bibnamefont{Bennett}} \bibnamefont{and}
  \bibinfo{author}{\bibfnamefont{G.}~\bibnamefont{Brassard}}, in
  \emph{\bibinfo{booktitle}{Proceedings of IEEE International Conference on
  Computers, Systems, and Signal Processing}} (\bibinfo{publisher}{IEEE},
  \bibinfo{year}{1984}), pp. \bibinfo{pages}{175--179};
\bibinfo{author}{\bibfnamefont{J.}~\bibnamefont{Barrett}},
  \bibinfo{author}{\bibfnamefont{L.}~\bibnamefont{Hardy}}, \bibnamefont{and}
  \bibinfo{author}{\bibfnamefont{A.}~\bibnamefont{Kent}},
  \bibinfo{journal}{Phys. Rev. Lett.} \textbf{\bibinfo{volume}{95}},
  \bibinfo{pages}{010503} (\bibinfo{year}{2005});
\bibinfo{author}{\bibfnamefont{A.}~\bibnamefont{Ac{\'\i}n}},
  \bibinfo{author}{\bibfnamefont{N.}~\bibnamefont{Gisin}}, \bibnamefont{and}
  \bibinfo{author}{\bibfnamefont{Ll.}~\bibnamefont{Masanes}}, \bibinfo{note}{quant-ph/0510094}.

\bibitem[{\citenamefont{Cirel'son}(1980)}]{Tsirelson:80a}
\bibinfo{author}{\bibfnamefont{B.~S.} \bibnamefont{Cirel'son}},
  \bibinfo{journal}{Lett. Math. Phys.} \textbf{\bibinfo{volume}{4}},
  \bibinfo{pages}{93} (\bibinfo{year}{1980}).

\bibitem[{\citenamefont{Tsirelson}(1987)}]{Tsirelson:85b}
\bibinfo{author}{\bibfnamefont{B.~S.} \bibnamefont{Tsirelson}},
  \bibinfo{journal}{J. Soviet Math.} \textbf{\bibinfo{volume}{36}},
  \bibinfo{pages}{557} (\bibinfo{year}{1987});
\bibinfo{author}{\bibfnamefont{B.}~\bibnamefont{Tsirelson}},
  \bibinfo{journal}{Hadronic J. Suppl.} \textbf{\bibinfo{volume}{8}},
  \bibinfo{pages}{329} (\bibinfo{year}{1993});
\bibinfo{author}{\bibfnamefont{R.}~\bibnamefont{Cleve}},
  \bibinfo{author}{\bibfnamefont{P.}~\bibnamefont{H{\o}yer}},
  \bibinfo{author}{\bibfnamefont{B.}~\bibnamefont{Toner}}, \bibnamefont{and}
  \bibinfo{author}{\bibfnamefont{J.}~\bibnamefont{Watrous}}, in
  \emph{\bibinfo{booktitle}{Proceedings of the 19th IEEE Conference on
  Computational Complexity (CCC 2004)}}, pp.
  \bibinfo{pages}{236--249};
\bibinfo{author}{\bibfnamefont{H.}~\bibnamefont{Buhrman}} \bibnamefont{and}
  \bibinfo{author}{\bibfnamefont{S.}~\bibnamefont{Massar}}, \bibinfo{note}{quant-ph/0409066};
\bibinfo{author}{\bibfnamefont{S.}~\bibnamefont{Wehner}},
  \bibinfo{journal}{Phys. Rev. A} \textbf{\bibinfo{volume}{73}},
  \bibinfo{eid}{022110} (\bibinfo{year}{2006}).

\bibitem[{\citenamefont{Toner}(2006)}]{toner:_monog}
\bibinfo{author}{\bibfnamefont{B.~F.} \bibnamefont{Toner}}, \bibinfo{note}{quant-ph/0601172}.

\bibitem[{\citenamefont{Navascues et~al.}()\citenamefont{Navascues, Pironio,
  and Acin}}]{navascues:_bound}
\bibinfo{author}{\bibfnamefont{M.}~\bibnamefont{Navascues}},
  \bibinfo{author}{\bibfnamefont{S.}~\bibnamefont{Pironio}}, \bibnamefont{and}
  \bibinfo{author}{\bibfnamefont{A.}~\bibnamefont{Acin}},
  \bibinfo{note}{quant-ph/0607119}.

\bibitem[{\citenamefont{Popescu and Rohrlich}(1994)}]{Popescu:94a}
\bibinfo{author}{\bibfnamefont{S.}~\bibnamefont{Popescu}} \bibnamefont{and}
  \bibinfo{author}{\bibfnamefont{D.}~\bibnamefont{Rohrlich}},
  \bibinfo{journal}{Found. Phys.} \textbf{\bibinfo{volume}{24}},
  \bibinfo{pages}{379} (\bibinfo{year}{1994});
\bibinfo{author}{\bibfnamefont{L.~A.} \bibnamefont{Khalfin}} \bibnamefont{and}
  \bibinfo{author}{\bibfnamefont{B.~S.} \bibnamefont{Tsirelson}}, in
  \emph{\bibinfo{booktitle}{Symposium on the Foundations of Modern Physics}},
  edited by \bibinfo{editor}{\bibfnamefont{P.}~\bibnamefont{Lahti}}
  \bibnamefont{and}
  \bibinfo{editor}{\bibfnamefont{P.}~\bibnamefont{Mittelstaedt}}
  (\bibinfo{publisher}{World Scientific}, \bibinfo{address}{Singapore},
  \bibinfo{year}{1985}), pp. \bibinfo{pages}{441--460}.


\bibitem[{\citenamefont{{van Dam}}(2005)}]{dam05:_implaus_conseq_super_nonloc}
\bibinfo{author}{\bibfnamefont{W.}~\bibnamefont{{van Dam}}}, \bibinfo{note}{quant-ph/0501159}; 
\bibinfo{author}{\bibfnamefont{G.}~\bibnamefont{Brassard}},
  \bibinfo{author}{\bibfnamefont{H.}~\bibnamefont{Buhrman}},
  \bibinfo{author}{\bibfnamefont{N.}~\bibnamefont{Linden}},
  \bibinfo{author}{\bibfnamefont{A.~A.} \bibnamefont{Methot}},
  \bibinfo{author}{\bibfnamefont{A.}~\bibnamefont{Tapp}}, \bibnamefont{and}
  \bibinfo{author}{\bibfnamefont{F.}~\bibnamefont{Unger}},
  \bibinfo{journal}{Phys. Rev. Lett.} \textbf{\bibinfo{volume}{96}},
  \bibinfo{eid}{250401} (\bibinfo{year}{2006}).

\bibitem[{\citenamefont{Coffman et~al.}(2000)\citenamefont{Coffman, Kundu, and
  Wootters}}]{Coffman:00a}
\bibinfo{author}{\bibfnamefont{V.}~\bibnamefont{Coffman}},
  \bibinfo{author}{\bibfnamefont{J.}~\bibnamefont{Kundu}}, \bibnamefont{and}
  \bibinfo{author}{\bibfnamefont{W.~K.} \bibnamefont{Wootters}},
  \bibinfo{journal}{Phys. Rev. A} \textbf{\bibinfo{volume}{61}},
  \bibinfo{pages}{052306} (\bibinfo{year}{2000});
\bibinfo{author}{\bibfnamefont{T.~J.} \bibnamefont{Osborne}} \bibnamefont{and}
  \bibinfo{author}{\bibfnamefont{F.}~\bibnamefont{Verstraete}},
  \bibinfo{journal}{Phys. Rev. Lett.} \textbf{\bibinfo{volume}{96}},
  \bibinfo{eid}{220503} (\bibinfo{year}{2006}).

\bibitem[{\citenamefont{Masanes et~al.}(2006)\citenamefont{Masanes, Ac{\'\i}n,
  and Gisin}}]{masanes05:_gener_nonsig_theor}
\bibinfo{author}{\bibfnamefont{Ll.}~\bibnamefont{Masanes}},
  \bibinfo{author}{\bibfnamefont{A.}~\bibnamefont{Ac{\'\i}n}},
  \bibnamefont{and} \bibinfo{author}{\bibfnamefont{N.}~\bibnamefont{Gisin}},
  \bibinfo{journal}{Phys. Rev. A} \textbf{\bibinfo{volume}{73}},
  \bibinfo{eid}{012112} (\bibinfo{year}{2006}).

\bibitem[{\citenamefont{Masanes}(2005)}]{masanes05:_extrem}
\bibinfo{author}{\bibfnamefont{Ll.}~\bibnamefont{Masanes}}, \bibinfo{note}{quant-ph/0512100}.

\bibitem[{\citenamefont{Bhatia}(1996)}]{bhatia96:_matrix_analy}
\bibinfo{author}{\bibfnamefont{R.}~\bibnamefont{Bhatia}},
  \emph{\bibinfo{title}{Matrix Analysis}}, vol. \bibinfo{volume}{169} of
  \emph{\bibinfo{series}{Graduate texts in mathematics}}
  (\bibinfo{publisher}{Springer-Verlag}, \bibinfo{address}{New York},
  \bibinfo{year}{1996}), \bibinfo{note}{section VII.1}.

\bibitem[{\citenamefont{Horodecki et~al.}(1996)\citenamefont{Horodecki,
  Horodecki, and Horodecki}}]{horodecki96:_separ}
\bibinfo{author}{\bibfnamefont{M.}~\bibnamefont{Horodecki}},
  \bibinfo{author}{\bibfnamefont{P.}~\bibnamefont{Horodecki}},
  \bibnamefont{and}
  \bibinfo{author}{\bibfnamefont{R.}~\bibnamefont{Horodecki}},
  \bibinfo{journal}{Phys. Lett. A} \textbf{\bibinfo{volume}{223}},
  \bibinfo{pages}{1} (\bibinfo{year}{1996}).



\end{thebibliography}

\end{document}